\pdfoutput=1

\documentclass[12pt,a4paper,twocolumn]{article}

\usepackage{ifthen} 
\newboolean{pdflatex}
\setboolean{pdflatex}{true} 

\newboolean{articletitles}
\setboolean{articletitles}{true} 

\newboolean{uprightparticles}
\setboolean{uprightparticles}{false} 

\newboolean{inbibliography}
\setboolean{inbibliography}{false} 

\usepackage{microtype}
\usepackage{lineno}  
\usepackage{xspace} 

\usepackage{graphicx}  
\usepackage{color}
\usepackage{colortbl}
\graphicspath{{./figs/}} 

\usepackage{amsmath} 
\usepackage{amssymb}
\usepackage{amsfonts}
\usepackage{upgreek} 

\newcommand*\patchAmsMathEnvironmentForLineno[1]{%
\expandafter\let\csname old#1\expandafter\endcsname\csname #1\endcsname
\expandafter\let\csname oldend#1\expandafter\endcsname\csname
end#1\endcsname
 \renewenvironment{#1}%
   {\linenomath\csname old#1\endcsname}%
   {\csname oldend#1\endcsname\endlinenomath}%
}
\newcommand*\patchBothAmsMathEnvironmentsForLineno[1]{%
  \patchAmsMathEnvironmentForLineno{#1}%
  \patchAmsMathEnvironmentForLineno{#1*}%
}
\AtBeginDocument{%
\patchBothAmsMathEnvironmentsForLineno{equation}%
\patchBothAmsMathEnvironmentsForLineno{align}%
\patchBothAmsMathEnvironmentsForLineno{flalign}%
\patchBothAmsMathEnvironmentsForLineno{alignat}%
\patchBothAmsMathEnvironmentsForLineno{gather}%
\patchBothAmsMathEnvironmentsForLineno{multline}%
}

\usepackage{hyperref}    
\usepackage[all]{hypcap} 

\def\lhcb {\mbox{LHCb}\xspace}

\ifthenelse{\boolean{uprightparticles}}%
{

 \def\Pmu         {\ensuremath{\upmu}\xspace}

 \def\Ppi         {\ensuremath{\uppi}\xspace}

 \def\Ppsi        {\ensuremath{\uppsi}\xspace}

 \def\PDelta      {\ensuremath{\Delta}\xspace}                 
 \def\PXi      {\ensuremath{\Xi}\xspace}                 
 \def\PLambda      {\ensuremath{\Lambda}\xspace}                 
 \def\PSigma      {\ensuremath{\Sigma}\xspace}                 
 \def\POmega      {\ensuremath{\Omega}\xspace}                 
 \def\PUpsilon      {\ensuremath{\Upsilon}\xspace}                 
 
 \def\PB      {\ensuremath{\mathrm{B}}\xspace}                 
                  
 \def\PD      {\ensuremath{\mathrm{D}}\xspace}

 \def\PJ      {\ensuremath{\mathrm{J}}\xspace}                 
 \def\PK      {\ensuremath{\mathrm{K}}\xspace}

 \def\PY      {\ensuremath{\mathrm{Y}}\xspace}

 \def\Pb      {\ensuremath{\mathrm{b}}\xspace}                 
 \def\Pc      {\ensuremath{\mathrm{c}}\xspace}                 
                  
 \def\Pe      {\ensuremath{\mathrm{e}}\xspace}

 \def\Pi      {\ensuremath{\mathrm{i}}\xspace}

 \def\Ps      {\ensuremath{\mathrm{s}}\xspace}

}
{

 \def\Pmu         {\ensuremath{\mu}\xspace}

 \def\Ppi         {\ensuremath{\pi}\xspace}

 \def\Ppsi        {\ensuremath{\psi}\xspace}                 
                  
 \mathchardef\PDelta="7101
 \mathchardef\PXi="7104
 \mathchardef\PLambda="7103
 \mathchardef\PSigma="7106
 \mathchardef\POmega="710A
 \mathchardef\PUpsilon="7107
                  
 \def\PB      {\ensuremath{B}\xspace}                 
                  
 \def\PD      {\ensuremath{D}\xspace}

 \def\PJ      {\ensuremath{J}\xspace}                 
 \def\PK      {\ensuremath{K}\xspace}

 \def\PY      {\ensuremath{Y}\xspace}

 \def\Pb      {\ensuremath{b}\xspace}                 
 \def\Pc      {\ensuremath{c}\xspace}                 
                  
 \def\Pe      {\ensuremath{e}\xspace}

 \def\Pi      {\ensuremath{i}\xspace}

 \def\Ps      {\ensuremath{s}\xspace}

}

\def\en         {\ensuremath{\Pe^-}\xspace}   
\def\ep         {\ensuremath{\Pe^+}\xspace}
 
\def\epem       {\ensuremath{\Pe^+\Pe^-}\xspace}

\def\mup        {\ensuremath{\Pmu^+}\xspace}
\def\mun        {\ensuremath{\Pmu^-}\xspace} 
\def\mumu       {\ensuremath{\Pmu^+\Pmu^-}\xspace}

\def\squark    {\ensuremath{\Ps}\xspace}

\def\cquark    {\ensuremath{\Pc}\xspace}
\def\cquarkbar {\ensuremath{\overline \cquark}\xspace}
\def\ccbar     {\ensuremath{\cquark\cquarkbar}\xspace}
\def\bquark    {\ensuremath{\Pb}\xspace}

\def\pion  {\ensuremath{\Ppi}\xspace}

\def\pip   {\ensuremath{\pion^+}\xspace}
\def\pim   {\ensuremath{\pion^-}\xspace}

\def\kaon  {\ensuremath{\PK}\xspace}
  \def\Kbar  {\kern 0.2em\overline{\kern -0.2em \PK}{}\xspace}

\def\Kp    {\ensuremath{\kaon^+}\xspace}

  \def\Dbar    {\kern 0.2em\overline{\kern -0.2em \PD}{}\xspace}
\def\D       {\ensuremath{\PD}\xspace}

\def\Dz      {\ensuremath{\D^0}\xspace}
\def\Dzb     {\ensuremath{\Dbar^0}\xspace}

\def\B       {\ensuremath{\PB}\xspace}
\def\Bbar    {\ensuremath{\kern 0.18em\overline{\kern -0.18em \PB}{}}\xspace}

\def\Bu      {\ensuremath{\B^+}\xspace}

\def\Bp      {\ensuremath{\Bu}\xspace}

\def\jpsi     {\ensuremath{{\PJ\mskip -3mu/\mskip -2mu\Ppsi\mskip 2mu}}\xspace}
\def\psitwos  {\ensuremath{\Ppsi{(2S)}}\xspace}
\def\psiprpr  {\ensuremath{\Ppsi(3770)}\xspace}

  \def\Y#1S{\ensuremath{\PUpsilon{(#1S)}}\xspace}

\def\Lbar {\ensuremath{\kern 0.1em\overline{\kern -0.1em\PLambda}}\xspace}

\def\BF         {{\ensuremath{\cal B}\xspace}}

\newcommand{\decay}[2]{\ensuremath{#1\!\to #2}\xspace}         

\def\to                 {\ensuremath{\rightarrow}\xspace}

\def\AT#1     {\ensuremath{A_{\mathrm{T}}^{#1}}\xspace}           

\def\C#1      {\ensuremath{\mathcal{C}_{#1}}\xspace}                       
\def\Cp#1     {\ensuremath{\mathcal{C}_{#1}^{'}}\xspace}                    
\def\Ceff#1   {\ensuremath{\mathcal{C}_{#1}^{\mathrm{(eff)}}}\xspace}        
\def\Cpeff#1  {\ensuremath{\mathcal{C}_{#1}^{'\mathrm{(eff)}}}\xspace}       
\def\Ope#1    {\ensuremath{\mathcal{O}_{#1}}\xspace}                       
\def\Opep#1   {\ensuremath{\mathcal{O}_{#1}^{'}}\xspace}                    

\newcommand{\tev}{\ifthenelse{\boolean{inbibliography}}{\ensuremath{~T\kern -0.05em eV}\xspace}{\ensuremath{\mathrm{\,Te\kern -0.1em V}}\xspace}}
\newcommand{\gev}{\ensuremath{\mathrm{\,Ge\kern -0.1em V}}\xspace}
\newcommand{\mev}{\ensuremath{\mathrm{\,Me\kern -0.1em V}}\xspace}
\newcommand{\kev}{\ensuremath{\mathrm{\,ke\kern -0.1em V}}\xspace}
\newcommand{\ev}{\ensuremath{\mathrm{\,e\kern -0.1em V}}\xspace}
\newcommand{\gevc}{\ensuremath{{\mathrm{\,Ge\kern -0.1em V\!/}c}}\xspace}
\newcommand{\mevc}{\ensuremath{{\mathrm{\,Me\kern -0.1em V\!/}c}}\xspace}
\newcommand{\gevcc}{\ensuremath{{\mathrm{\,Ge\kern -0.1em V\!/}c^2}}\xspace}
\newcommand{\gevgevcccc}{\ensuremath{{\mathrm{\,Ge\kern -0.1em V^2\!/}c^4}}\xspace}
\newcommand{\mevcc}{\ensuremath{{\mathrm{\,Me\kern -0.1em V\!/}c^2}}\xspace}

\def\mum  {\ensuremath{\,\upmu\rm m}\xspace}

\def\invfb   {\ensuremath{\mbox{\,fb}^{-1}}\xspace}

\newcommand{\chisq}{\ensuremath{\chi^2}\xspace}

\newcommand{\chisqip}{\ensuremath{\chi^2_{\rm IP}}\xspace}

\def\gsim{{~\raise.15em\hbox{$>$}\kern-.85em
          \lower.35em\hbox{$\sim$}~}\xspace}
\def\lsim{{~\raise.15em\hbox{$<$}\kern-.85em
          \lower.35em\hbox{$\sim$}~}\xspace}

\def\pt         {\mbox{$p_{\rm T}$}\xspace}

\def\mrad{\ensuremath{\rm \,mrad}\xspace}

\def\gauss      {\mbox{\textsc{Gauss}}\xspace}

\def\tell1  {TELL1\xspace}
\def\ukl1   {UKL1\xspace}

\usepackage{cite} 
\usepackage{mciteplus}

\def\qq {\ensuremath{q^{2}}\xspace}

\def\KMuMu     {\ensuremath{\Kp\mup\mun}\xspace}
\def\BuJpsiK       {\decay{\Bp}{ \jpsi \Kp}}
\def\BuKMuMu     {\decay{\Bp}{ \Kp \mup\mun}}
\def\BupsitwosK       {\decay{\Bp}{ \psitwos \Kp}}

\def\dimuon {\ensuremath{m_{\mup\mun}}\xspace}

\usepackage{sectsty}
\sectionfont{\large}

\begin{document}

\renewcommand{\thefootnote}{\fnsymbol{footnote}}
\setcounter{footnote}{1}

\onecolumn
\begin{titlepage}
\pagenumbering{roman}

\vspace*{-1.5cm}
\centerline{\large EUROPEAN ORGANIZATION FOR NUCLEAR RESEARCH (CERN)}
\vspace*{1.5cm}
\hspace*{-0.5cm}
\begin{tabular*}{\linewidth}{lc@{\extracolsep{\fill}}r}
\ifthenelse{\boolean{pdflatex}}
{\vspace*{-2.7cm}\mbox{\!\!\!\includegraphics[width=.14\textwidth]{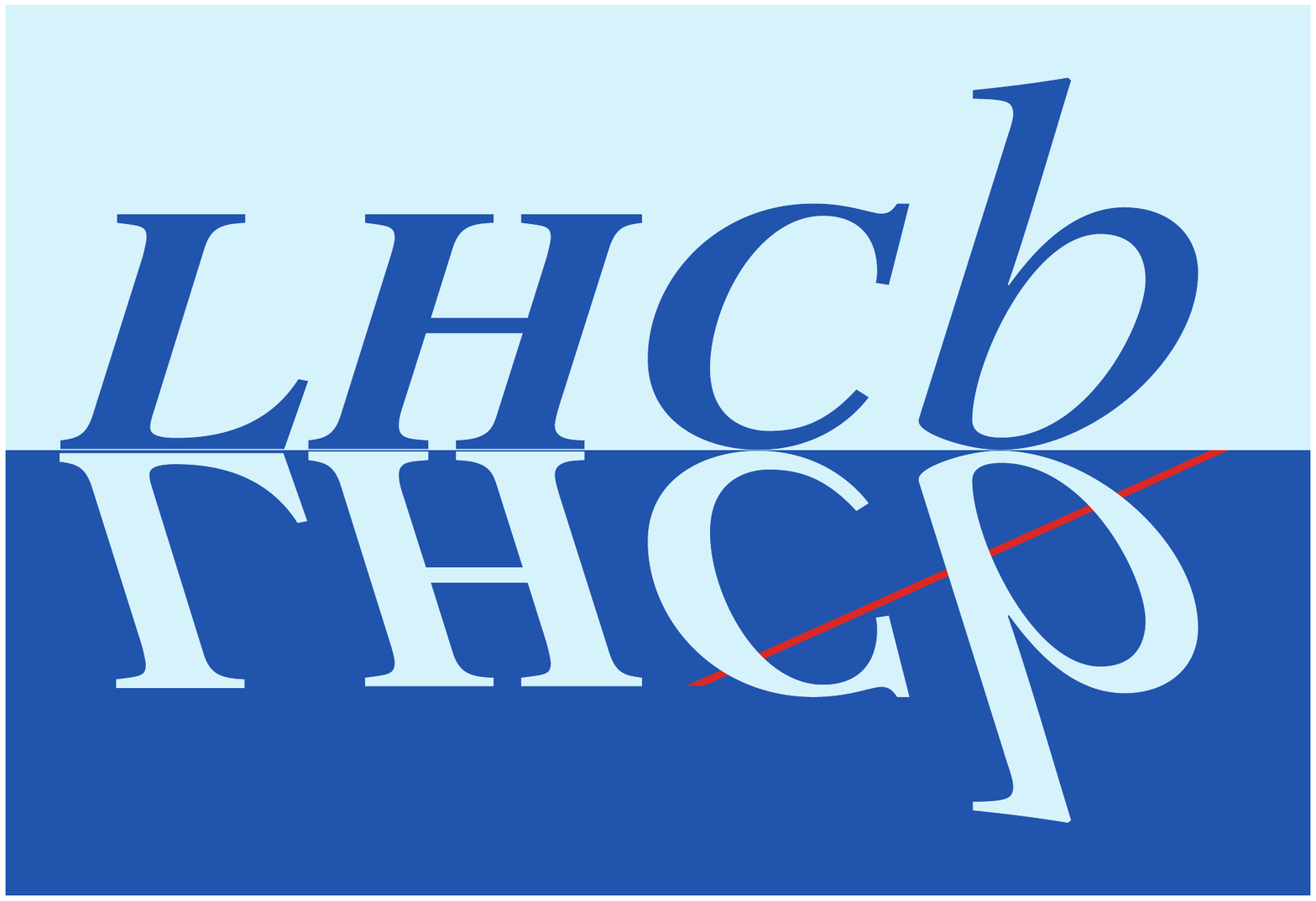}} & &}%
{\vspace*{-1.2cm}\mbox{\!\!\!\includegraphics[width=.12\textwidth]{lhcb-logo.eps}} & &}%
\\
 & & CERN-PH-EP-2013-137 \\  
 & & LHCb-PAPER-2013-039 \\  
 & & 4 September 2013 \\ 
 & & \\
\end{tabular*}

\vspace*{3.4cm}

{\bf\boldmath\huge
\begin{center}
  Observation of a resonance in $B^+ \to K^+ \mu^+\mu^-$
  decays at low recoil
\end{center}
}

\vspace*{1.8cm}

\begin{center}
The LHCb collaboration\footnote{Authors are listed on the following pages.}
\end{center}

\vspace{\fill}

\begin{abstract}
  \noindent
A broad peaking structure is observed in the dimuon spectrum of $B^+
\to K^+ \mu^+\mu^-$ decays in the kinematic region where the kaon has
a low recoil against the dimuon system. The structure is consistent
with interference between the $B^+ \to K^+ \mu^+\mu^-$ decay and a resonance and has a
statistical significance exceeding six standard deviations. The mean
and width of the resonance are measured to be $4191^{+9}_{-8}\mathrm{\,Me\kern
-0.1em V\!/}c^2$ and $65^{+22}_{-16}\mathrm{\,Me\kern -0.1em
V\!/}c^2$, respectively, where the uncertainties include statistical and systematic contributions. 
These measurements are compatible with the properties of the $\psi(4160)$ meson. First
observations of both the decay $B^+ \to \psi(4160) K^+$ and the
subsequent decay $\psi(4160) \to \mu^+\mu^-$ are reported. The
resonant decay and the interference contribution make up 20\,\% of the yield for dimuon 
masses above 3770\mevcc. This contribution is larger than theoretical estimates.
  
\end{abstract}
\vspace*{1.8cm}

\begin{center}
  Accepted by Phys.~Rev.~Lett. 
\end{center}

\vspace{\fill}

{\footnotesize 
\centerline{\copyright~CERN on behalf of the \lhcb collaboration, license \href{http://creativecommons.org/licenses/by/3.0/}{CC-BY-3.0}.}}
\vspace*{2mm}

\end{titlepage}


\newpage
\setcounter{page}{2}
\mbox{~}
\newpage

\centerline{\large\bf LHCb collaboration}
\begin{flushleft}
\small
R.~Aaij$^{40}$, 
B.~Adeva$^{36}$, 
M.~Adinolfi$^{45}$, 
C.~Adrover$^{6}$, 
A.~Affolder$^{51}$, 
Z.~Ajaltouni$^{5}$, 
J.~Albrecht$^{9}$, 
F.~Alessio$^{37}$, 
M.~Alexander$^{50}$, 
S.~Ali$^{40}$
% ...

{\footnotesize \it
% ...
$ ^{5}$Clermont Universit\'{e}, Universit\'{e} Blaise Pascal, CNRS/IN2P3, LPC, Clermont-Ferrand, France\\
$ ^{6}$CPPM, Aix-Marseille Universit\'{e}, CNRS/IN2P3, Marseille, France\\
% ...
$ ^{9}$Fakult\"{a}t Physik, Technische Universit\"{a}t Dortmund, Dortmund, Germany\\
% ...
$ ^{36}$Universidad de Santiago de Compostela, Santiago de Compostela, Spain\\
$ ^{37}$European Organization for Nuclear Research (CERN), Geneva, Switzerland\\
% ...
$ ^{40}$Nikhef National Institute for Subatomic Physics, Amsterdam, The Netherlands\\
% ...
$ ^{45}$H.H. Wills Physics Laboratory, University of Bristol, Bristol, United Kingdom\\
% ...
$ ^{50}$School of Physics and Astronomy, University of Glasgow, Glasgow, United Kingdom\\
$ ^{51}$Oliver Lodge Laboratory, University of Liverpool, Liverpool, United Kingdom\\
% ...
}
\end{flushleft}

\cleardoublepage

\twocolumn

\renewcommand{\thefootnote}{\arabic{footnote}}
\setcounter{footnote}{0}

\pagestyle{plain} 
\setcounter{page}{1}
\pagenumbering{arabic}


\noindent The decay of the \Bp meson to the final state \KMuMu
receives contributions from tree level decays and decays mediated
through virtual quantum loop processes. The tree level decays proceed
through the decay of a \Bp meson to a vector \ccbar resonance and a \Kp meson,
followed by the decay of the resonance to a pair of muons. Decays
mediated by flavour changing neutral current~(FCNC) loop processes
give rise to pairs of muons with a non-resonant mass distribution. To
probe contributions to the FCNC decay from physics beyond the Standard Model (SM),
 it is essential that the tree level decays are properly
accounted for. In all analyses of the \BuKMuMu decay, from
discovery~\cite{Abe:2001dh} to the latest most accurate
measurement~\cite{LHCb-PAPER-2012-024}, this has been done by placing a
veto on the regions of dimuon mass, \dimuon, dominated by the \jpsi and
\psitwos resonances. In the low recoil region, corresponding to a dimuon mass above 
the open charm threshold, theoretical predictions of the decay rate can
be obtained with an operator product expansion
(OPE)~\cite{Grinstein:2004vb} in which the \ccbar contribution
and other hadronic effects are treated as effective
interactions.

Nearly all available information about the ${ J^{PC}}=1^{--}$ charmonium
resonances above the open charm threshold, where the resonances are
wide as decays to $\PD$$^{(*)}\overline{\PD}{^{(*)}}$ are allowed, comes from
measurements of the cross-section ratio of $\epem \to {\rm hadrons}$
relative to $\epem \to \mumu$. Among these analyses,
only that of the BES collaboration in Ref.~\cite{Ablikim:2007gd}
takes interference and strong phase differences between the different
resonances into account. The broad and overlapping nature of these 
resonances means that they cannot be excluded by vetoes on the 
dimuon mass in an efficient way, and a more sophisticated treatment is required. 

This Letter describes a measurement of a broad peaking structure in
the low recoil region of the \decay{\Bp}{\Kp\mumu} decay, based on
data corresponding to an integrated luminosity of 3\invfb taken with
the \lhcb detector at a centre-of-mass energy of $7\tev$ in 2011 and $8\tev$ in 2012. 
Fits to the dimuon mass spectrum are performed, where one or several resonances are
allowed to interfere with the non-resonant \BuKMuMu signal, and their
parameters determined. The inclusion of charge conjugated processes is implied throughout this Letter. 

The \lhcb detector~\cite{Alves:2008zz} is a single-arm forward
spectrometer covering the \mbox{pseudorapidity} range $2<\eta <5$,
designed for the study of particles containing \bquark or \cquark
quarks. The detector includes a high-precision tracking system
consisting of a silicon-strip vertex detector surrounding the $pp$
interaction region, a large-area silicon-strip detector located
upstream of a dipole magnet with a bending power of about
$4{\rm\,Tm}$, and three stations of silicon-strip detectors and straw
drift tubes placed downstream.  The combined tracking system provides
a momentum measurement with relative uncertainty that varies from
0.4\,\% at 5\gevc to 0.6\,\% at 100\gevc, and impact parameter
resolution of 20\mum for tracks with high transverse momentum. Charged
hadrons are identified using two ring-imaging Cherenkov detectors.
Muons are identified by a system composed of alternating layers of
iron and multiwire proportional chambers. Simulated events used 
in this analysis are produced using the software
described in Refs.~\cite{Sjostrand:2006za,LHCb-PROC-2010-056,
Lange:2001uf,Golonka:2005pn,Allison:2006ve, *Agostinelli:2002hh,LHCb-PROC-2011-006}.

Candidates are required to pass a two stage trigger system~\cite{LHCb-DP-2012-004}. 
In the initial hardware stage, candidate events are selected with at least one muon with
transverse momentum, $\pt>1.48\,(1.76)\gevc$ in 2011~(2012). In the
subsequent software stage, at least one of the final state particles
is required to have both $\pt> 1.0\gevc$ and 
impact parameter larger than $100\mum$ with respect to all of the primary $pp$
interaction vertices~(PVs) in the event. Finally, a multivariate
algorithm~\cite{BBDT} is used for the identification of secondary
vertices consistent with the decay of a \bquark hadron with muons in
the final state.

The selection of the \KMuMu final state is made in two steps. Candidates
are required to pass an initial selection, which reduces the data
sample to a manageable level, followed by a multivariate
selection. The dominant background is of a combinatorial
nature, where two correctly identified muons from different heavy
flavour hadron decays are combined with a kaon from either of those
decays. This category of background has no peaking structure in either
the dimuon mass or the \KMuMu mass. The signal region is defined as
 $5240 < m_{\Kp\mumu} < 5320 \mevcc$ and the
sideband region as $5350 < m_{\Kp\mumu} < 5500 \mevcc$. The sideband
below the \Bp mass is not used as it contains backgrounds from
partially reconstructed decays, which do not contaminate the
signal region.

The initial selection requires: $\chisqip>9$ for all final state
particles, where $\chisqip$ is defined as the minimum change in
$\chisq$ when the particle is included in a vertex fit to any of the
PVs in the event; that the muons are
positively identified in the muon system; and that the dimuon vertex has a
vertex fit $\chisq<9$. In addition, based on the
lowest $\chisqip$ of the \Bp candidate, an associated PV is
chosen. For this PV it is required that: the \Bp candidate has $\chisqip<16$; the
vertex fit $\chi^{2}$ must increase by more than $121$
when including the \Bp candidate daughters; and the angle between 
the \Bp candidate momentum and the direction from the PV
 to the decay vertex should be below 14\mrad. Finally, the \Bp candidate is required to
have a vertex fit $\chisq<24$ (with three degrees of freedom).

The multivariate selection is based on a boosted decision
tree~(BDT)~\cite{Breiman} with the Ada\-Boost algorithm\cite{AdaBoost}
to separate signal from background. It is trained with a signal sample
from simulation and a background sample consisting of 10\,\% of the
data from the sideband region. The
multivariate selection uses geometric and kinematic variables, where
the most discriminating variables are the \chisqip of the final state
particles and the vertex quality of the \Bp candidate. The selection
with the BDT has an efficiency of 90\,\% on signal surviving the initial selection while retaining
6\,\% of the background. The overall efficiency for the
reconstruction, trigger and selection, normalised to the total number of \BuKMuMu decays produced at
the LHCb interaction point, is $2\,\%$. As the branching fraction measurements are normalised
to the \BuJpsiK decay, only relative efficiencies are used. The yields in the \KMuMu final state from
\BuJpsiK and \BupsitwosK decays are $9.6\times10^{5}$ and $8\times10^{4}$ events,
respectively.

In addition to the combinatorial background, there
are several small sources of potential background that form a peak in either or
both of the $m_{\KMuMu}$ and \dimuon distributions. The largest of
these backgrounds are the decays \BuJpsiK and \BupsitwosK, where 
the kaon and one of the muons have been interchanged. 
The decays \decay{\Bp}{\Kp\pim\pip} and \decay{\Bp}{\Dzb \pip} followed by
\decay{\Dzb}{\Kp\pim}, with the two pions identified as muons are also considered. 
To reduce these backgrounds to a negligible level, tight particle
identification criteria and vetoes on 
$\mun\Kp$ combinations compatible with \jpsi, \psitwos, or
\Dz meson decays are applied. 
These vetoes are 99\% efficient on signal.

A kinematic fit~\cite{Hulsbergen:2005pu}
is performed for all selected candidates. In the fit the \KMuMu mass is constrained to the nominal
\Bp mass and the candidate is required to originate from its associated PV. For \BupsitwosK
decays, this improves the resolution in \dimuon from 15\mevcc to
5\mevcc. Given the widths of the resonances that are
subsequently analysed, resolution effects are neglected. 
While the \psitwos state is narrow, the large 
branching fraction means that its non-Gaussian tail is significant and hard to model.
The \psitwos contamination is reduced to a negligible level by requiring 
$\dimuon > 3770\mevcc$. This dimuon mass range is defined as the low recoil region used in this analysis.

In order to estimate the amount of background present in the \dimuon
spectrum, an unbinned extended maximum likelihood fit is performed to
the \KMuMu mass distribution without the \Bp mass constraint. The signal shape is taken from a mass
fit to the \BupsitwosK mode in data with the shape parameterised as the sum of
two Crystal Ball functions~\cite{Skwarnicki:1986xj}, with common tail
parameters, but different widths. The Gaussian width of the two
components is increased by $5\,\%$ for the fit to the low recoil
region as determined from simulation. The low recoil region contains 1830 candidates in the signal mass
window, with a signal to background ratio of 7.8. 

\begin{figure}[tb]
  \centering
  \includegraphics[width=0.98\linewidth]{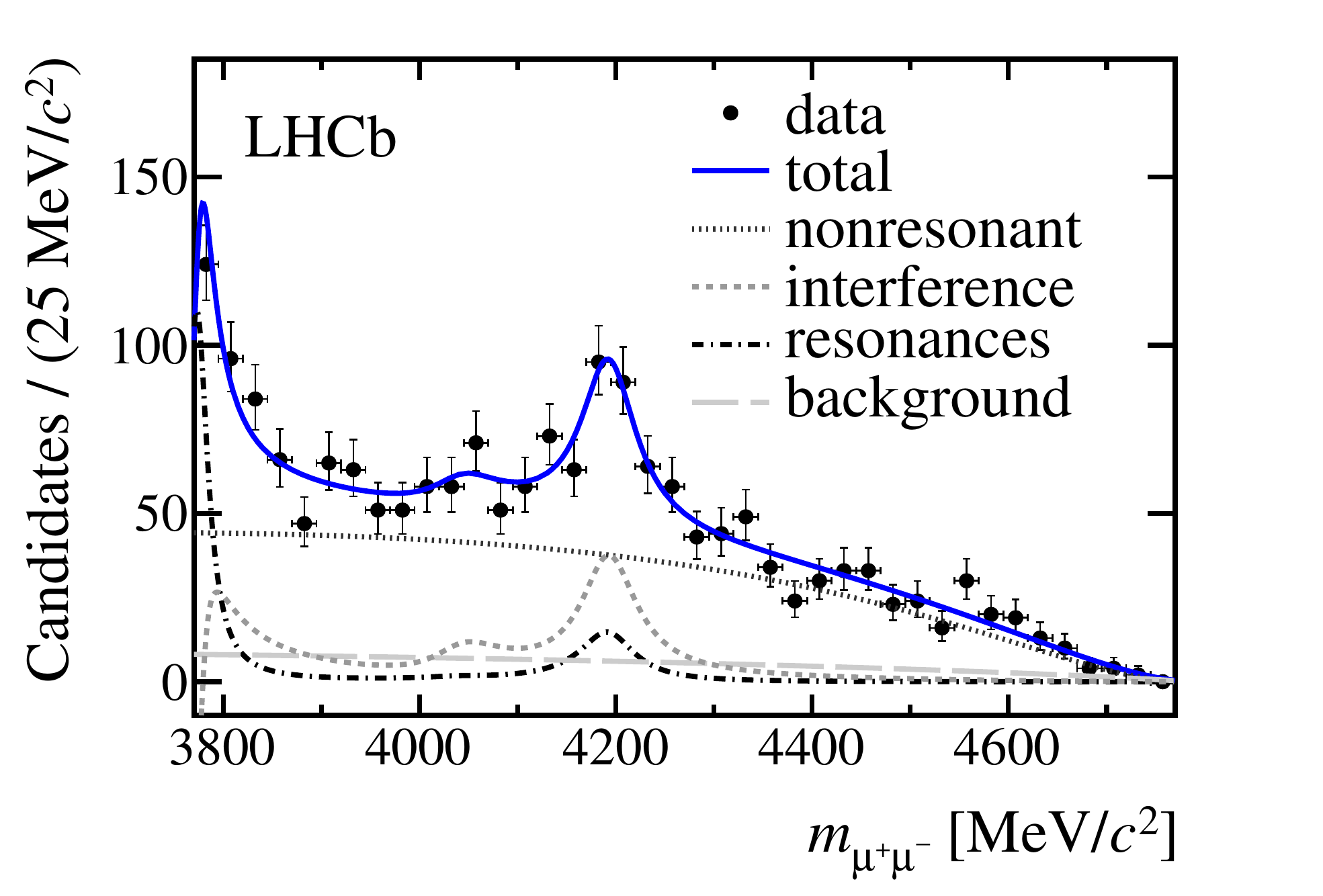}  
  \caption{\small Dimuon mass distribution of data with fit results
    overlaid for the fit that includes contributions from the
    non-resonant vector and axial vector components, and the \psiprpr,
    $\psi(4040)$, and $\psi(4160)$ resonances. Interference terms are
    included and the relative strong phases are left free in the fit.}
  \label{fig:fitresults}
\end{figure}

The dimuon mass distribution in the low recoil region is shown in
Fig.~\ref{fig:fitresults}. Two peaks are visible, one at the
low edge corresponding to the expected decay $\decay{\psiprpr}{\mumu}$ and a
wide peak at a higher mass. In all fits, a vector resonance component corresponding 
to this decay is included. Several fits are made to the distribution.
The first introduces a vector resonance with unknown
parameters. Subsequent fits look at the compatibility of the data
with the hypothesis that the peaking structure is due to known
resonances.

The non-resonant part of the mass fits contains a vector and
axial vector component. Of these, only the vector component will
interfere with the resonance. The probability
density function~(PDF) of the signal component is given as
\begin{align}
{\cal P}_{\rm sig} &\propto P(\dimuon) \ |{\cal A}|^{2} \ f^{2}(m_{\mumu}^{2}) \, ,
\label{eq:Psig} \\
 |{\cal A}|^{2} & = |A^{\rm{V}}_{\text{nr}} + 
     \sum_{k} e^{i\delta_{k}}A_{\text{r}}^{k}|^{2} + |A^{\rm{AV}}_{\text{nr}}|^{2} \, ,
\label{eq:Asum}
\end{align}
where $A^{\rm{V}}_{\text{nr}}$ and $A^{\rm{AV}}_{\text{nr}}$ are the
vector and axial vector amplitudes of the non-resonant decay. The shape
of the non-resonant signal in \dimuon is driven by phase space, $P(m_{\mumu})$, and
the form factor, $f(m_{\mumu}^{2})$. The parametrisation of
Ref.~\cite{Khodjamirian:2010vf} is used to describe the dimuon mass
dependence of the form factor. This form factor parametrisation is 
consistent with recent lattice
calculations~\cite{lattice}. In the SM at low recoil, the ratio of the vector and 
axial vector contributions to the non-resonant component is expected to have negligible 
dependence on the dimuon mass. The vector component accounts for
 $(45\pm 6)\, \%$ of the differential branching fraction in the SM 
(see, for example, Ref.~\cite{Bobeth:2011nj}). This estimate of the 
vector component is assumed in the fit.

The total vector amplitude is formed by summing the vector amplitude of
the non-resonant signal with a number of Breit-Wigner amplitudes,
$A_{\text{r}}^{k}$, which depend on $m_{\mup\mun}$. Each Breit-Wigner amplitude is rotated by a
phase, $\delta_k$, which represents the strong phase difference
between the non-resonant vector component and the resonance with index
$k$. Such phase differences are
expected~\cite{Khodjamirian:2010vf}. The \psiprpr resonance, visible at the lower
edge of the dimuon mass distribution, is included in the fit
as a Breit-Wigner component whose mass and width are constrained to
the world average values~\cite{PDG2012}. 

The background PDF for the dimuon
mass distribution is taken from a fit to data in the \KMuMu sideband.
The uncertainties on the background amount and shape are included as
Gaussian constraints to the fit in the signal region.  

The signal PDF
is multiplied by the relative efficiency as a function of dimuon mass
with respect to the \BuJpsiK decay. As in previous analyses of the
same final state~\cite{LHCb-PAPER-2012-011}, this efficiency is
determined from simulation after the simulation is made to match data
by: degrading by $\sim\!20\,\%$ the impact parameter resolution of the
tracks, reweighting events to match the kinematic properties of the
\Bp candidates and the track multiplicity of the event, and
adjusting the particle identification variables based on calibration
samples from data. In the region from the \jpsi mass to $4600\mevcc$
the relative efficiency drops by around 20\,\%. From there to the
kinematic endpoint it drops sharply, predominantly due to the \chisqip
cut on the kaon as in this region its direction is aligned with the 
\Bp candidate and therefore also with the PV.

Initially, a fit with a single resonance in addition to the \psiprpr and non-resonant terms is
performed. This additional resonance has its phase, mean, and width left
free. The parameters of the resonance returned by the fit are a mass
of $4191^{+9}_{-8}\mevcc$ and a width of
$65^{+22}_{-16}\mevcc$. Branching fractions are determined by integrating the square of the
Breit-Wigner amplitude returned by the fit, normalising to the \BuJpsiK yield, 
and multiplying with the product of branching fractions,
$\BF(\BuJpsiK)\times\BF(\decay{\jpsi}{\mumu})$~\cite{PDG2012}. The
product $\BF(\decay{\Bp}{X\Kp})\times\BF(\decay{X}{\mumu})$ 
for the additional resonance, $X$, is determined to be
$(3.9^{+0.7}_{-0.6}) \times 10^{-9}$. The uncertainty on this product is
calculated using the profile likelihood. The data are not sensitive to the vector 
fraction of the non-resonant component as the branching fraction of the 
resonance will vary to compensate. For example, if the vector fraction 
is lowered to $30\%$, the central value of the branching fraction increases
 to $4.6\times10^{-9}$. This reflects the lower amount of interference 
 allowed between the resonant and non-resonant components.
 
The significance of the resonance is obtained
by simulating pseudo-experiments that include the non-resonant,
\psiprpr and background components. The log likelihood ratios
between fits that include and exclude a resonant component for $6
\times 10^5$ such samples are compared to the difference observed in
fits to the data. None of the samples have a higher ratio than observed
in data and an extrapolation gives a significance of the signal above
six standard deviations.

The properties of the resonance are compatible with the mass and width of the
$\psi(4160)$ resonance as measured in Ref.~\cite{Ablikim:2007gd}. 
To test the hypothesis that $\psi$ resonances well above the open 
charm threshold are observed, another fit including the
$\psi(4040)$ and $\psi(4160)$ resonances is performed. The mass and width of
the two are constrained to the measurements from
Ref.~\cite{Ablikim:2007gd}. The data have no sensitivity to a
$\psi(4415)$ contribution. The fit describes the data well and the
parameters of the $\psi(4160)$ meson are almost unchanged with respect to
the unconstrained fit. The fit overlaid on the data is shown in
Fig.~\ref{fig:fitresults} and Table~\ref{tab:results} reports the fit
parameters.
\begin{table}
  \centering
  \caption{\small Parameters of the dominant resonance 
  for fits where the mass and width are unconstrained and
  constrained to those of the $\psi(4160)$ meson~\cite{Ablikim:2007gd}, respectively. The branching 
    fractions are for the \Bp decay followed by the decay of the 
    resonance to muons.}
  \setlength{\extrarowheight}{2pt}
  \begin{tabular}{l r r}
    \hline
                          & \hspace{-0.6cm} Unconstrained & $\psi(4160)$ \\
    \hline
    \BF $[\times 10^{-9}$] & $3.9\,^{+0.7}_{-0.6}$ & $3.5\,^{+0.9}_{-0.8}$ \\
    Mass $[\mevcc]$       & $4191\,^{+9}_{-8}$   &  $4190 \pm 5$ \\
    Width $[\mevcc]$      & $65\,^{+22}_{-16}$    & $66 \pm 12$ \\
    Phase [rad]           & $-1.7\pm0.3$       &  $-1.8\pm0.3$ \\
    \hline
  \end{tabular}
  \label{tab:results}
\end{table}

\begin{figure}[b]
  \centering
  \includegraphics[width=0.98\linewidth]{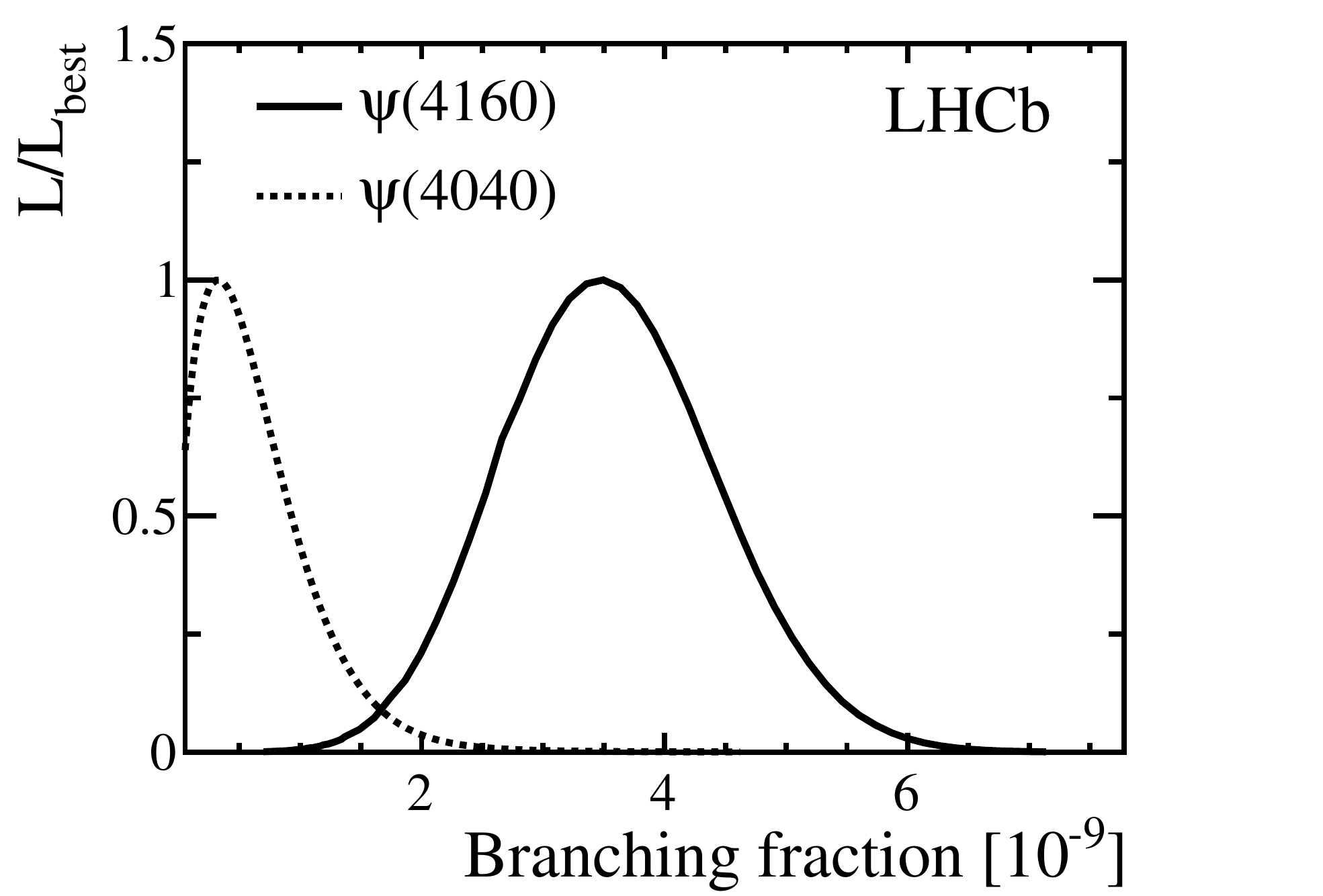}
  \caption{\small Profile likelihood ratios for the product of branching
    fractions
    $\BF(\decay{\Bp}{\psi\Kp})\times\BF(\decay{\psi}{\mumu})$ of the
    $\psi(4040)$ and the $\psi(4160)$ mesons. At each point all other fit parameters are reoptimised.}
  \label{fig:BFscan}
\end{figure}

The resulting profile likelihood ratio compared to the best
fit as a function of branching fraction can be seen in
Fig.~\ref{fig:BFscan}. In the fit with the three $\psi$ resonances, the
$\psi(4160)$ meson is visible with
$\BF(\decay{\Bp}{\psi(4160)\Kp})\times\BF(\decay{\psi(4160)}{\mumu}) =
(3.5^{+0.9}_{-0.8}) \times 10^{-9}$ but for the $\psi(4040)$ meson, no significant signal is seen, and an upper limit is set. The limit
$\BF(\decay{\Bp}{\psi(4040)\Kp})\times\BF(\decay{\psi(4040)}{\mumu}) <
1.3\,(1.5) \times 10^{-9}$ at $90\,(95)\,\%$ confidence level is obtained by integrating
the likelihood ratio compared to the best fit and assuming a flat
prior for any positive branching fraction.

\begin{figure}[tb]
  \centering
  \includegraphics[width=0.98\linewidth]{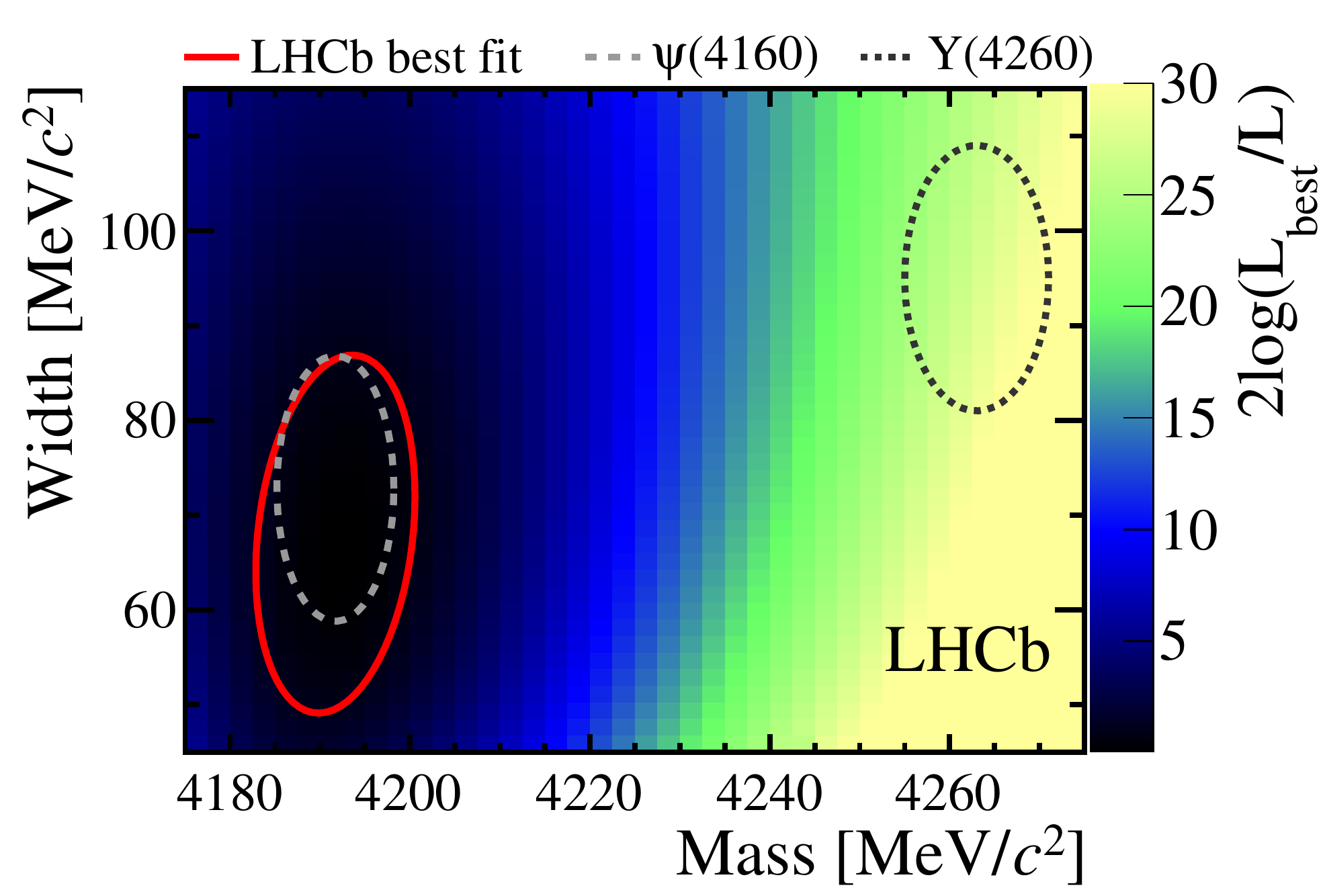}
  \caption{\small Profile likelihood as a function of mass and
    width of a fit with a single extra resonance. At each point all
    other fit parameters are reoptimised. The three ellipses
    are (red-solid) the best fit and previous measurements of (grey-dashed) the
    $\psi(4160)$~\cite{Ablikim:2007gd} and (black-dotted) the
    $\PY(4260)$~\cite{PDG2012} states.}
  \label{fig:2dscan}
\end{figure}

In Fig.~\ref{fig:2dscan} the likelihood scan of the fit with a single
extra resonance is shown as a function of the mass and width of the
resonance. The fit is compatible with the
$\psi(4160)$ resonance, while a hypothesis where the resonance corresponds to the decay
$\decay{\PY(4260)}{\mumu}$ is disfavoured by more than four standard
deviations.

Systematic uncertainties associated with the normalisation procedure are 
negligible as the decay \BuJpsiK has the same final state as the signal and 
similar kinematics. Uncertainties due to the resolution and mass scale are insignificant. The systematic
uncertainty associated to the form factor parameterisation in the fit
model is taken from Ref.~\cite{Bobeth:2011nj}. Finally, the uncertainty on the vector
fraction of the non-resonant amplitude is obtained using the EOS
tool described in Ref.~\cite{Bobeth:2011nj} and is dominated
by the uncertainty from short distance contributions. All systematic uncertainties 
are included in the fit as Gaussian constraints. From comparing
the difference in the uncertainties on masses, widths and branching
fractions for fits with and without these systematic constraints, it can
be seen that the systematic uncertainties are about 20\,\% the size of
the statistical uncertainties and thus contribute less than 2\% to the total uncertainty.

In summary, a resonance has been observed in the dimuon spectrum of \BuKMuMu
decays with a significance of above six standard deviations. The resonance can be
explained by the contribution of the $\psi(4160)$, via the decays
\decay{\Bp}{\psi(4160)\Kp} and \decay{\psi(4160)}{\mumu}. It
constitutes first observations of both decays. The $\psi(4160)$ is
known to decay to electrons with a branching fraction of
$(6.9\pm4.0)\times10^{-6}$~\cite{Ablikim:2007gd}. Assuming lepton
universality, the branching fraction of the decay
\decay{\Bp}{\psi(4160)\Kp} is measured to be
$(5.1^{+1.3}_{-1.2}\pm3.0)\times10^{-4}$, where the second uncertainty
corresponds to the uncertainty on the \decay{\psi(4160)}{\ep\en}
branching fraction. The corresponding limit for \decay{\Bp}{\psi(4040)\Kp} is
calculated to be $1.3\,(1.7)\times10^{-4}$ at a 90\,(95)\,\% confidence level. The absence of the decay
\decay{\Bp}{\psi(4040)\Kp} at a similar level is interesting, and
suggests future studies of \BuKMuMu decays based on larger 
datasets may reveal new insights into \ccbar spectroscopy.

The contribution of the $\psi(4160)$ resonance in the low recoil region, taking into
account interference with the non-resonant \decay{\Bp}{\Kp\mumu} decay, is about 20\,\% of the total signal. 
This value is larger than theoretical estimates, where the \ccbar contribution is $\sim$10\% of the 
vector amplitude, with a small correction from quark-hadron duality violation~\cite{Beylich:2011aq}. 
Results presented in this Letter will play an important role in controlling charmonium effects 
in future inclusive and exclusive \mbox{$\bquark\to\squark\mumu$} measurements.

\section*{Acknowledgements}
\noindent We express our gratitude to our colleagues in the CERN
accelerator departments for the excellent performance of the LHC. We
thank the technical and administrative staff at the LHCb
institutes. We acknowledge support from CERN and from the national
agencies: CAPES, CNPq, FAPERJ and FINEP (Brazil); NSFC (China);
CNRS/IN2P3 and Region Auvergne (France); BMBF, DFG, HGF and MPG
(Germany); SFI (Ireland); INFN (Italy); FOM and NWO (The Netherlands);
SCSR (Poland); MEN/IFA (Romania); MinES, Rosatom, RFBR and NRC
``Kurchatov Institute'' (Russia); MinECo, XuntaGal and GENCAT (Spain);
SNSF and SER (Switzerland); NAS Ukraine (Ukraine); STFC (United
Kingdom); NSF (USA). We also acknowledge the support received from the
ERC under FP7. The Tier1 computing centres are supported by IN2P3
(France), KIT and BMBF (Germany), INFN (Italy), NWO and SURF (The
Netherlands), PIC (Spain), GridPP (United Kingdom). We are thankful
for the computing resources put at our disposal by Yandex LLC
(Russia), as well as to the communities behind the multiple open
source software packages that we depend on.

\addcontentsline{toc}{section}{References}
\setboolean{inbibliography}{true}
\ifx\mcitethebibliography\mciteundefinedmacro
\PackageError{LHCb.bst}{mciteplus.sty has not been loaded}
{This bibstyle requires the use of the mciteplus package.}\fi
\providecommand{\href}[2]{#2}

\newpage
\onecolumn
\appendix


\begin{mcitethebibliography}{10}
\mciteSetBstSublistMode{n}
\mciteSetBstMaxWidthForm{subitem}{\alph{mcitesubitemcount})}
\mciteSetBstSublistLabelBeginEnd{\mcitemaxwidthsubitemform\space}
{\relax}{\relax}

\bibitem{Abe:2001dh}
Belle collaboration, K.~Abe {\em et~al.},
  \ifthenelse{\boolean{articletitles}}{{\it {Observation of the decay $B \to K
  \ell^{+} \ell^{-}$}},
  }{}\href{http://dx.doi.org/10.1103/PhysRevLett.88.021801}{Phys.\ Rev.\ Lett.\
   {\bf 88} (2001) 021801}, \href{http://arxiv.org/abs/hep-ex/0109026}{{\tt
  arXiv:hep-ex/0109026}}\relax
\mciteBstWouldAddEndPuncttrue
\mciteSetBstMidEndSepPunct{\mcitedefaultmidpunct}
{\mcitedefaultendpunct}{\mcitedefaultseppunct}\relax
\EndOfBibitem
\bibitem{LHCb-PAPER-2012-024}
LHCb collaboration, R.~Aaij {\em et~al.},
  \ifthenelse{\boolean{articletitles}}{{\it {Differential branching fraction
  and angular analysis of the $B^{+} \rightarrow K^{+}\mu^{+}\mu^{-}$ decay}},
  }{}\href{http://dx.doi.org/10.1007/JHEP02(2013)105}{JHEP {\bf 02} (2013)
  105}, \href{http://arxiv.org/abs/1209.4284}{{\tt arXiv:1209.4284}}\relax
\mciteBstWouldAddEndPuncttrue
\mciteSetBstMidEndSepPunct{\mcitedefaultmidpunct}
{\mcitedefaultendpunct}{\mcitedefaultseppunct}\relax
\EndOfBibitem
\bibitem{Grinstein:2004vb}
B.~Grinstein and D.~Pirjol, \ifthenelse{\boolean{articletitles}}{{\it
  {Exclusive rare $B \to K^* \ell^+\ell^-$ decays at low recoil: controlling
  the long-distance effects}},
  }{}\href{http://dx.doi.org/10.1103/PhysRevD.70.114005}{Phys.\ \, Rev.\  {\bf
  D70} (2004) 114005}, \href{http://arxiv.org/abs/hep-ph/0404250}{{\tt
  arXiv:hep-ph/0404250}}\relax
\mciteBstWouldAddEndPuncttrue
\mciteSetBstMidEndSepPunct{\mcitedefaultmidpunct}
{\mcitedefaultendpunct}{\mcitedefaultseppunct}\relax
\EndOfBibitem
\bibitem{Ablikim:2007gd}
BES collaboration, M.~Ablikim {\em et~al.},
  \ifthenelse{\boolean{articletitles}}{{\it {Determination of the $\psi(3770)$,
  $\psi(4040)$, $\psi(4160)$ and $\psi(4415)$ resonance parameters}},
  }{}\href{http://dx.doi.org/10.1016/j.physletb.2007.11.100}{Phys.\ Lett {\bf
  B660} (2008) 315}, \href{http://arxiv.org/abs/0705.4500}{{\tt
  arXiv:0705.4500}}\relax
\mciteBstWouldAddEndPuncttrue
\mciteSetBstMidEndSepPunct{\mcitedefaultmidpunct}
{\mcitedefaultendpunct}{\mcitedefaultseppunct}\relax
\EndOfBibitem
\bibitem{Alves:2008zz}
LHCb collaboration, A.~A. Alves~Jr. {\em et~al.},
  \ifthenelse{\boolean{articletitles}}{{\it {The \lhcb detector at the LHC}},
  }{}\href{http://dx.doi.org/10.1088/1748-0221/3/08/S08005}{JINST {\bf 3}
  (2008) S08005}\relax
\mciteBstWouldAddEndPuncttrue
\mciteSetBstMidEndSepPunct{\mcitedefaultmidpunct}
{\mcitedefaultendpunct}{\mcitedefaultseppunct}\relax
\EndOfBibitem
\bibitem{Sjostrand:2006za}
T.~Sj\"{o}strand, S.~Mrenna, and P.~Skands,
  \ifthenelse{\boolean{articletitles}}{{\it {PYTHIA 6.4 physics and manual}},
  }{}\href{http://dx.doi.org/10.1088/1126-6708/2006/05/026}{JHEP {\bf 05}
  (2006) 026}, \href{http://arxiv.org/abs/hep-ph/0603175}{{\tt
  arXiv:hep-ph/0603175}}\relax
\mciteBstWouldAddEndPuncttrue
\mciteSetBstMidEndSepPunct{\mcitedefaultmidpunct}
{\mcitedefaultendpunct}{\mcitedefaultseppunct}\relax
\EndOfBibitem
\bibitem{LHCb-PROC-2010-056}
I.~Belyaev {\em et~al.}, \ifthenelse{\boolean{articletitles}}{{\it {Handling of
  the generation of primary events in \gauss, the \lhcb simulation framework}},
  }{}\href{http://dx.doi.org/10.1109/NSSMIC.2010.5873949}{Nuclear Science
  Symposium Conference Record (NSS/MIC) {\bf IEEE} (2010) 1155}\relax
\mciteBstWouldAddEndPuncttrue
\mciteSetBstMidEndSepPunct{\mcitedefaultmidpunct}
{\mcitedefaultendpunct}{\mcitedefaultseppunct}\relax
\EndOfBibitem
\bibitem{Lange:2001uf}
D.~J. Lange, \ifthenelse{\boolean{articletitles}}{{\it {The EvtGen particle
  decay simulation package}},
  }{}\href{http://dx.doi.org/10.1016/S0168-9002(01)00089-4}{Nucl.\ Instrum.\
  Meth.\  {\bf A462} (2001) 152}\relax
\mciteBstWouldAddEndPuncttrue
\mciteSetBstMidEndSepPunct{\mcitedefaultmidpunct}
{\mcitedefaultendpunct}{\mcitedefaultseppunct}\relax
\EndOfBibitem
\bibitem{Golonka:2005pn}
P.~Golonka and Z.~Was, \ifthenelse{\boolean{articletitles}}{{\it {PHOTOS Monte
  Carlo: a precision tool for QED corrections in $Z$ and $W$ decays}},
  }{}\href{http://dx.doi.org/10.1140/epjc/s2005-02396-4}{Eur.\ Phys.\ J.\  {\bf
  C45} (2006) 97}, \href{http://arxiv.org/abs/hep-ph/0506026}{{\tt
  arXiv:hep-ph/0506026}}\relax
\mciteBstWouldAddEndPuncttrue
\mciteSetBstMidEndSepPunct{\mcitedefaultmidpunct}
{\mcitedefaultendpunct}{\mcitedefaultseppunct}\relax
\EndOfBibitem
\bibitem{Allison:2006ve}
Geant4 collaboration, J.~Allison {\em et~al.},
  \ifthenelse{\boolean{articletitles}}{{\it Geant4 developments and
  applications}, }{}\href{http://dx.doi.org/10.1109/TNS.2006.869826}{IEEE
  Trans.\ Nucl.\ Sci.\  {\bf 53} (2006) 270}\relax
\mciteBstWouldAddEndPuncttrue
\mciteSetBstMidEndSepPunct{\mcitedefaultmidpunct}
{\mcitedefaultendpunct}{\mcitedefaultseppunct}\relax
\EndOfBibitem
\bibitem{Agostinelli:2002hh}
Geant4 collaboration, S.~Agostinelli {\em et~al.},
  \ifthenelse{\boolean{articletitles}}{{\it {Geant4: a simulation toolkit}},
  }{}\href{http://dx.doi.org/10.1016/S0168-9002(03)01368-8}{Nucl.\ Instrum.\
  Meth.\  {\bf A506} (2003) 250}\relax
\mciteBstWouldAddEndPuncttrue
\mciteSetBstMidEndSepPunct{\mcitedefaultmidpunct}
{\mcitedefaultendpunct}{\mcitedefaultseppunct}\relax
\EndOfBibitem
\bibitem{LHCb-PROC-2011-006}
M.~Clemencic {\em et~al.}, \ifthenelse{\boolean{articletitles}}{{\it {The \lhcb
  simulation application, \gauss: design, evolution and experience}},
  }{}\href{http://dx.doi.org/10.1088/1742-6596/331/3/032023}{{J.\ Phys.\ \!\!:
  Conf.\ Ser.\ } {\bf 331} (2011) 032023}\relax
\mciteBstWouldAddEndPuncttrue
\mciteSetBstMidEndSepPunct{\mcitedefaultmidpunct}
{\mcitedefaultendpunct}{\mcitedefaultseppunct}\relax
\EndOfBibitem
\bibitem{LHCb-DP-2012-004}
R.~Aaij {\em et~al.}, \ifthenelse{\boolean{articletitles}}{{\it {The \lhcb
  trigger and its performance in 2011}},
  }{}\href{http://dx.doi.org/10.1088/1748-0221/8/04/P04022}{JINST {\bf 8}
  (2013) P04022}, \href{http://arxiv.org/abs/1211.3055}{{\tt
  arXiv:1211.3055}}\relax
\mciteBstWouldAddEndPuncttrue
\mciteSetBstMidEndSepPunct{\mcitedefaultmidpunct}
{\mcitedefaultendpunct}{\mcitedefaultseppunct}\relax
\EndOfBibitem
\bibitem{BBDT}
V.~V. Gligorov and M.~Williams, \ifthenelse{\boolean{articletitles}}{{\it
  {Efficient, reliable and fast high-level triggering using a bonsai boosted
  decision tree}},
  }{}\href{http://dx.doi.org/10.1088/1748-0221/8/02/P02013}{JINST {\bf 8}
  (2013) P02013}, \href{http://arxiv.org/abs/1210.6861}{{\tt
  arXiv:1210.6861}}\relax
\mciteBstWouldAddEndPuncttrue
\mciteSetBstMidEndSepPunct{\mcitedefaultmidpunct}
{\mcitedefaultendpunct}{\mcitedefaultseppunct}\relax
\EndOfBibitem
\bibitem{Breiman}
L.~Breiman, J.~H. Friedman, R.~A. Olshen, and C.~J. Stone, {\em Classification
  and regression trees}, Wadsworth international group, Belmont, California,
  USA, 1984\relax
\mciteBstWouldAddEndPuncttrue
\mciteSetBstMidEndSepPunct{\mcitedefaultmidpunct}
{\mcitedefaultendpunct}{\mcitedefaultseppunct}\relax
\EndOfBibitem
\bibitem{AdaBoost}
R.~E. Schapire and Y.~Freund, \ifthenelse{\boolean{articletitles}}{{\it A
  decision-theoretic generalization of on-line learning and an application to
  boosting}, }{}\href{http://dx.doi.org/10.1006/jcss.1997.1504}{Jour.\ Comp.\
  and Syst.\ Sc.\  {\bf 55} (1997) 119}\relax
\mciteBstWouldAddEndPuncttrue
\mciteSetBstMidEndSepPunct{\mcitedefaultmidpunct}
{\mcitedefaultendpunct}{\mcitedefaultseppunct}\relax
\EndOfBibitem
\bibitem{Hulsbergen:2005pu}
W.~D. Hulsbergen, \ifthenelse{\boolean{articletitles}}{{\it {Decay chain
  fitting with a Kalman filter}},
  }{}\href{http://dx.doi.org/10.1016/j.nima.2005.06.078}{Nucl.\ Instrum.\
  Meth.\  {\bf A552} (2005) 566},
  \href{http://arxiv.org/abs/physics/0503191}{{\tt
  arXiv:physics/0503191}}\relax
\mciteBstWouldAddEndPuncttrue
\mciteSetBstMidEndSepPunct{\mcitedefaultmidpunct}
{\mcitedefaultendpunct}{\mcitedefaultseppunct}\relax
\EndOfBibitem
\bibitem{Skwarnicki:1986xj}
T.~Skwarnicki, {\em A study of the radiative cascade transitions between the
  Upsilon-prime and Upsilon resonances}, PhD thesis, Institute of Nuclear
  Physics, Krakow, 1986,
  {\href{http://inspirehep.net/record/230779/files/230779.pdf}{DESY-F31-86-02}}\relax
\mciteBstWouldAddEndPuncttrue
\mciteSetBstMidEndSepPunct{\mcitedefaultmidpunct}
{\mcitedefaultendpunct}{\mcitedefaultseppunct}\relax
\EndOfBibitem
\bibitem{Khodjamirian:2010vf}
A.~Khodjamirian, {Th. Mannel}, A.~A. Pivovarov, and Y.-M. Wang,
  \ifthenelse{\boolean{articletitles}}{{\it {Charm-loop effect in $B \to
  K^{(*)} \ell^{+} \ell^{-}$ and $B\to K^*\gamma$}},
  }{}\href{http://dx.doi.org/10.1007/JHEP09(2010)089}{JHEP {\bf 09} (2010)
  089}, \href{http://arxiv.org/abs/1006.4945}{{\tt arXiv:1006.4945}}\relax
\mciteBstWouldAddEndPuncttrue
\mciteSetBstMidEndSepPunct{\mcitedefaultmidpunct}
{\mcitedefaultendpunct}{\mcitedefaultseppunct}\relax
\EndOfBibitem
\bibitem{lattice}
C.~Bouchard {\em et~al.}, \ifthenelse{\boolean{articletitles}}{{\it {The rare
  decay $B \rightarrow K \ell^{+} \ell^{-}$ form factors from lattice QCD}},
  }{}
\newblock \href{http://arxiv.org/abs/1306.2384}{{\tt arXiv:1306.2384}}\relax
\mciteBstWouldAddEndPuncttrue
\mciteSetBstMidEndSepPunct{\mcitedefaultmidpunct}
{\mcitedefaultendpunct}{\mcitedefaultseppunct}\relax
\EndOfBibitem
\bibitem{Bobeth:2011nj}
C.~Bobeth, G.~Hiller, D.~van Dyk, and C.~Wacker,
  \ifthenelse{\boolean{articletitles}}{{\it {The decay $B \to K
  \ell^{+}\ell^{-}$ at low hadronic recoil and model-independent $\Delta B = 1$
  constraints}}, }{}\href{http://dx.doi.org/10.1007/JHEP01(2012)107}{JHEP {\bf
  01} (2012) 107}, \href{http://arxiv.org/abs/1111.2558}{{\tt
  arXiv:1111.2558}}\relax
\mciteBstWouldAddEndPuncttrue
\mciteSetBstMidEndSepPunct{\mcitedefaultmidpunct}
{\mcitedefaultendpunct}{\mcitedefaultseppunct}\relax
\EndOfBibitem
\bibitem{PDG2012}
Particle Data Group, J.~Beringer {\em et~al.},
  \ifthenelse{\boolean{articletitles}}{{\it {\href{http://pdg.lbl.gov/}{Review
  of particle physics}}},
  }{}\href{http://dx.doi.org/10.1103/PhysRevD.86.010001}{Phys.\ Rev.\  {\bf
  D86} (2012) 010001}\relax
\mciteBstWouldAddEndPuncttrue
\mciteSetBstMidEndSepPunct{\mcitedefaultmidpunct}
{\mcitedefaultendpunct}{\mcitedefaultseppunct}\relax
\EndOfBibitem
\bibitem{LHCb-PAPER-2012-011}
LHCb collaboration, R.~Aaij {\em et~al.},
  \ifthenelse{\boolean{articletitles}}{{\it {Measurement of the isospin
  asymmetry in $B \to K^{(*)}\mu^ + \mu^-$ decays}},
  }{}\href{http://dx.doi.org/10.1007/JHEP07(2012)133}{JHEP {\bf 07} (2012)
  133}, \href{http://arxiv.org/abs/1205.3422}{{\tt arXiv:1205.3422}}\relax
\mciteBstWouldAddEndPuncttrue
\mciteSetBstMidEndSepPunct{\mcitedefaultmidpunct}
{\mcitedefaultendpunct}{\mcitedefaultseppunct}\relax
\EndOfBibitem
\bibitem{Beylich:2011aq}
M.~Beylich, G.~Buchalla, and T.~Feldmann,
  \ifthenelse{\boolean{articletitles}}{{\it {Theory of $\B \to K^{(*)} \ell^{+}
  \ell^{-}$ decays at high \qq: OPE and quark-hadron duality}},
  }{}\href{http://dx.doi.org/10.1140/epjc/s10052-011-1635-0}{Eur.\ Phys.\ J.\
  {\bf C71} (2011) 1635}, \href{http://arxiv.org/abs/1101.5118}{{\tt
  arXiv:1101.5118}}\relax
\mciteBstWouldAddEndPuncttrue
\mciteSetBstMidEndSepPunct{\mcitedefaultmidpunct}
{\mcitedefaultendpunct}{\mcitedefaultseppunct}\relax
\EndOfBibitem
\end{mcitethebibliography}
\end{document}